\documentstyle[prl,aps,multicol,epsf]{revtex}
\begin{document}
\draft

\title{
Generic mechanism for generating a liquid-liquid phase transition.
}
\author{
Giancarlo Franzese$^1$, Gianpietro Malescio$^2$, Anna Skibinsky$^1$,
Sergey V. Buldyrev$^1$, and H. Eugene Stanley$^1$ 
}

\address{
$^1$Center for Polymer Studies and Department of Physics,
Boston Univ., Boston, MA 02215, USA\\
$^2$Dipartimento di Fisica, Universit\`a di Messina 
and Istituto Nazionale Fisica della Materia, 98166 Messina, Italy
}
\date{\today}

\maketitle

\begin{multicols}{2}

{\bf Recent experimental results \cite{Katayama} indicate that
phosphorus, a single-component system, can have two liquid phases: a
high-density liquid (HDL) and a low-density liquid (LDL) phase.  
A
first-order transition between two liquids of different densities
\cite{debenedetti} is consistent with
experimental data for a variety
of materials \cite{Wilding,Brazhkin97},
including single-component systems such as water
\cite{Brazhkin98,mishima_2000,bellisent,soper}, silica
\cite{Lacks} and carbon \cite{Thiel}.  Molecular dynamics simulations of
very specific models for supercooled water \cite{debenedetti,Poole},
liquid carbon \cite{Glosli} and supercooled silica \cite{Saika-Voivod},
predict a LDL-HDL critical point, but a coherent and general
interpretation of the LDL-HDL transition is lacking.  Here we show that
the presence of a LDL and a HDL can be directly related to an
interaction potential with an attractive part and 
two characteristic short-range repulsive distances.  
This kind of
interaction is common to other single-component materials in the liquid
state (in particular liquid metals
\cite{Mon,hs1,Silbert,Levesque,Kincaid,Cummings,Velasco,Voronel,debenedetti}),
and such potentials are often used to decribe systems that exhibit a
density anomaly \cite{debenedetti}. However, our results 
show  that the LDL and HDL phases can occur in systems with no
density anomaly.
Our results therefore present  an experimental challenge
to uncover a liquid-liquid transition in  systems like liquid metals,
regardless of the presence of the density anomaly}.

Several explanations have been developed to understand the liquid-liquid
phase transition. For example, the
{\it two-liquid models} \cite{Brazhkin97} assume that 
liquids at high pressure are a mixture of 
 two liquid phases  whose relative 
concentration depends on external parameters.
Other explanations for 
the liquid-liquid phase transition assume an anisotropic potential 
\cite{debenedetti,Poole,Glosli,Saika-Voivod}.
Here we shall see that
liquid-liquid phase transition phenomena can arise solely from
 an isotropic pair interaction potential with two characteristic lengths.

For molecular liquid phosphorus P$_4$ (as for water), a tetrahedral open
structure is preferred at low pressures $P$ and low temperatures $T$,
while a denser structure is favored at high $P$ and high $T$
\cite{Katayama,mishima_2000,soper}.  The existence of these two structures
with different densities suggests a pair interaction with two
characteristic distances.  
The first distance can be associated with the {\it hard-core} exclusion
between two particles and the second distance with a weak repulsion ({\it
soft-core}), which can be overcome at large pressure. Here we will use a 
generic three dimensional (3D) model composed of particles interacting
via an isotropic soft-core pair potential.
Such isotropic potentials can be regarded as resulting from an average
over the angular part of more realistic potentials, and are often used
as a first approximation to understand the qualitative behavior of real
systems
\cite{Mon,hs1,Silbert,Levesque,Kincaid,Cummings,Velasco,Voronel,colloids,debenedetti}.  
For Ce and Cs, Stell and Hemmer proposed a potential with
nearest-neighbor 
repulsion and a weak long-range attraction \cite{hs1}. By means
of an exact analysis in 1D, they found two critical points,
with the high-density critical point interpreted as a
solid-solid transition.
Then analytic calculations \cite{Silbert}, simulations  \cite{Levesque}
and exact solution in 1D \cite{Kincaid} of the structure
factor for a model with a soft-core potential were found consistent with 
experimental structure factors for liquid metals such as Bi. 
The structure factor was also the focus
of a theoretical study of a family of soft-core
potentials, for liquid metals, by means of mean-spherical approximation
 \cite{Cummings}.
More recently, the analysis of the solid phase of a model with  a soft-core
potential \cite{Velasco} was related to the experimental evidence of
 a liquid-liquid critical point in the K$_2$Cs metallic alloy
\cite{Voronel}.  Moreover, for simple metals, soft-core potentials have
been derived by first-principle calculations \cite{Mon} and computed
from experimental data
\cite{debenedetti}.  The importance of soft-core potentials in the
context of supercooled liquids was pointed out by Debenedetti and others
\cite{DRB,ssbs,jagla}, and the analysis of experimental data for water
gives rise to a soft-core potential \cite{SH-G1}.

The isotropic pair potential considered here  (Fig.\ref{franzese_fig1}a, inset)
has a hard-core  radius $a$ and a
soft-core  radius $b>a$. For
$a<r<b$ the particles repel each other with energy $U_R>0$, for $b<r<c$
they attract each other with energy $-U_A<0$.
For $r>c$ the interaction
is considered negligible and is approximated to zero.
The potential has three free parameters: $b/a$, $c/a$ and $U_R/U_A$.

To select the parameters for the molecular dynamics (MD) simulations is
not easy, and MD is too time consuming to study a wide range of
parameter values.  Hence we first perform an
integral equation analysis
\cite{caccamo}, whose predictions we can calculate very rapidly and
efficiently. 
In this technique, one derives the phase diagram by studying
the static pair distribution function -- which 
measures the probability of finding a particle at a given 
distance from a reference particle, and thereby quantifies the
correlation between pairs of particles. Mathematically, this function can be
written as the sum of an infinite series of  
many-dimensional integrals, over particle coordinates, involving the pair
interaction potential.  Since this exact
expression is intractable, approximations must be made.
Here we use the hypernetted-chain approximation, which
consists of neglecting a
specific class of these integrals, leading to a simplified integral
equation that can be solved numerically. 
In the temperature - density ($T$-$\rho$) phase diagram, the region 
 where the simplified equation has no solutions is
related \cite{caccamo} to the region where the system separates in two fluid phases.
Thus this technique allows us to estimate the parameter range 
where two critical points occur, and hence
to find useful parameters values for the MD
simulations: $b/a=2.0$, $c/a=2.2$ and
$U_R/U_A=0.5$. We also use in the  MD calculations  several additional
parameter sets.

Specifically we  perform  MD simulations in 3D at constant volume $V$
and number of particles $N=490$ and 850.
We use periodic boundary conditions,
a standard collision
event list algorithm \cite{ssbs}, and a modified Berendsen method to control $T$
\protect\cite{b}. We find, for each set of parameters,  the appearance of two critical
points (Fig.\ref{franzese_fig1}).  

A critical point is revealed by the presence of a region, in the
$P$-$\rho$ phase diagram, with negative-slope isotherms.  In MD
simulations this region is related to the
coexistence of two phases \cite{debenedetti}. 
The (local) maximum and minimum along an
isotherm correspond to the limits of stability of the existence of each
single phase (supercooled and superheated phase, respectively).  By
definition, these maxima and minima are points on the {\it spinodal
line} for that temperature.  Since the spinodal line
has a maximum at a critical point, 
a way to locate a critical point is to find this maximum.
In our simulations (Fig.\ref{franzese_fig1}), we find two regions with
negatively-sloped isotherms and the overall shape of the spinodal line
has two maxima, showing the presence of {\it two} critical points, $C_1$
and $C_2$.  Using the Maxwell construction in the $P$-$V$ plane
\cite{debenedetti}, we evaluate the {\it coexistence lines} of the two
fluid phases associated with each critical point (Fig.\ref{franzese_fig2}).
Considering both the maxima of the spinodal line and the maxima of the
coexistence regions in the $P$-$\rho$ and $P$-$T$ planes, we
estimate the low-density critical point $C_1$ at $T_1=0.606\pm 0.004
~U_A/k_B$, $P_1=0.0177\pm 0.0008 ~U_A/a^{3}$, $\rho_1=0.11\pm 0.01
~a^{-3}$ and the high-density critical point $C_2$ at $T_2=0.665\pm
0.005 ~U_A/k_B$, $P_2=0.10\pm 0.01 ~U_A/a^{3}$, $\rho_2=0.32\pm 0.03
~a^{-3}$.  Critical point $C_1$ is at the end of the phase transition line
separating the gas phase and the LDL phase, while critical point $C_2$
is at the end of the phase transition line separating the gas phase and the
HDL phase.  Their relative positions resemble the phosphorus phase
diagram, except that, in the experiments, $C_2$ has not been located
\cite{Katayama}, but is expected at the end of the gas-HDL transition
line.

Our phase diagram (Fig.\ref{franzese_fig2}) shows the following fluid phases.  At
high $T>T_2$, the only fluid phase is the gas.  At $T_1< T < T_2$, we
find -- depending on $\rho$ -- the gas alone, or the HDL alone (turquoise
region), or the HDL coexisting with gas (black line in Fig.\ref{franzese_fig2}a and green
 region  in Fig.\ref{franzese_fig2}c).  Below $T_1$, the
LDL phase appears alone (blue region), or in coexistence with the HDL
(orange line in Fig.\ref{franzese_fig2}b and orange region in Fig.\ref{franzese_fig2}d), 
or in  
coexistence with the gas (red line in Fig.\ref{franzese_fig2}b and red region in
Fig.\ref{franzese_fig2}d).  The point where the gas-LDL coexistence line merges
with the LDL-HDL coexistence line is the {\it triple point}.  Below the
pressure and temperature of the triple point, LDL is not stable and
separates into gas and HDL.

For phosphorus
the liquid-liquid transition occurs in the stable
fluid regime \cite{Katayama}. In contrast, for our model, it occurs in
the metastable fluid regime (see Fig.1). We therefore whish 
to understand how to enhance the stability
of the critical points with respect to the crystal phase.
We find that 
by increasing the attractive well width $(c-b)/a$,
both critical temperatures $T_1$ and $T_2$ increase, and hence  both
critical points 
move toward  the stable fluid phase, analogous to results for
attractive potentials with  a single critical point \cite{Frenkel,Hagen}.
For example, 
for attractive well width
$(c-b)/a=0.2$, both $C_1$ and $C_2$ are {\it metastable} with
respect to the crystal, while for $(c-b)/a>0.7$
we find $C_1$ in the {\it stable} fluid phase.

The phase diagram 
depends sensitively also on the relative width of the shoulder 
$b/a$ and on its relative height $U_R/U_A$.  By
decreasing $b/a$ or by increasing $U_R/U_A$, $T_2$ decreases and becomes
smaller than $T_1$. This means that, in these cases, the high-density
$C_2$ occurs below the temperature of the gas-liquid critical point,
i.e. $C_2$ is in the liquid phase and represents a LDL-HDL critical
point, as in supercooled water \cite{mishima_2000,bellisent,Poole}.

The soft-core potential with the sets of parameter
we use displays no ``density anomaly''
$(\partial V/\partial T)_P<0$.
This result is at first sight surprising since soft-core potentials have
often been used to explain the density anomaly (see, e.g.,
Refs.\cite{debenedetti,ssbs}).
To understand this result, we consider the
entropy $S$ (the degree of disorder in the system) and  
the thermodynamic relation 
$-(\partial V/\partial T)_P = (\partial S/\partial P)_T = 
(\partial S/\partial V)_T (\partial V/\partial P)_T$.
Since of necessity $(\partial V/\partial P)_T<0$, 
$(\partial V/\partial T)_P<0$ implies $(\partial S/\partial V)_T<0$,
i.e. the density anomaly implies that the disorder in the system
increases for decreasing volume. For example, this is the case for
water. This is consistent with the negative slope $dP/dT$ of the
crystal-liquid transition line for water, that
implies $\Delta S/\Delta V<0$
for the 
Clausius-Clapeyron equation  $dP/dT=\Delta S/\Delta V$,
where $\Delta S$ and $\Delta V$ are the entropy
and volume differences between the two coexisting phases.

For our system, we expect the reverse: $(\partial V/\partial T)_P>0$ so 
$(\partial S/\partial V)_T>0$, consistent with the positive 
slope of the LDL-HDL transition
line $dP/dT$ (see Fig.\ref{franzese_fig2}b). 
We confirm our expectation that $(\partial S/\partial V)_T>0$ by explicitly
calculating $S$ for our system by means of thermodynamic integration.

Our results show 
that the presence of two critical points and the occurrence of the density
anomaly are not necessarily related, suggesting that one might seek
 experimental evidence of a liquid-liquid phase transition in systems
with no density anomaly.  In particular, a second critical point may
also exist in liquid metals that can be described by soft-core
potentials.  Thus the class
of experimental systems displaying a second critical point may be
broader than previously hypothesized.

{\bf Acknowledgments}

We wish to thank L.A.N. Amaral, P.V. Giaquinta, E. La Nave, T. Lopez Ciudad,
S. Mossa, G. Pellicane, A. Scala,
F.W. Starr, J. Teixeira, and, in particular, F. Sciortino and two
anonymous referees for helpful suggestions and discussions.  We thank
NSF and CNR
(Italy) for partial support.

\begin{figure}
\caption{ Pressure-density isotherms,
crystallization line and spinodal line from the MD simulations for the
isotropic pair potential in 3D.  
{\bf (a), Inset:} 
The pair potential energy $U(r)$  as a function of the distance $r$ between
two particles.
{\bf (a)} 
Several isotherms
for (bottom to top) $k_BT/U_A=0.57$, 0.59, 0.61, 0.63,
0.65, 0.67 ($k_B$ is the Boltzmann constant).  Diamonds represent data
points and lines are guides for the eyes. 
The solid line connecting  local
maxima and minima along the isotherms represents the {\it spinodal
line}.  The two maxima 
of the spinodal line (squares) represent the two critical
points $C_1$ and $C_2$. 
To determine the {\it crystallization line} (gray line) -- below which
the fluid is metastable with respect to the crystal -- we place a crystal
seed, prepared at very low $T$, in contact with the fluid, and check,
for each  ($T$,$\rho$), 
if the seed grows or melts after $10^6$ MD steps.
The spontaneous formation ({\it nucleation}) of the crystal is observed, within our simulation times
($\approx {10^5}$ MD steps), only for $\rho\geq 0.27~a^{-3}$.  
We use the structure factor $S(Q)$ -- the
Fourier transform of the
density-density correlation function for wave vectors $Q$ -- to
determine when the nucleation occurs. Indeed, 
at the onset of  nucleation,
$S(Q)$ develops large peaks at finite $Q$ ($Q=12 a^{-1}$ and $Q=6 a^{-1}$).  
For each $\rho$, we
quench the system from a high-$T$ configuration. After a transient time
for the fluid equilibration, we
compute $P(T,\rho)$, averaging over $10^5$ -- $10^6$
configurations generated from up to 12 independent quenches, making sure
that the calculations are done before nucleation takes place.  
{\bf (b)} 
Enlarged view of the region around the gas-LDL critical point $C_1$  for 
$k_BT/U_A= 0.570$, 0.580, 0.590, 0.595, 0.600, 0.610, 0.620, 0.630.}
\label{franzese_fig1}
\end{figure}

\begin{figure}
\caption{
The phase diagrams, with coexistence lines and critical points 
resulting from MD simulations.
{\bf (a,b)} 
$P$-$T$ phase diagram. Panel (b) is a blow up of panel (a) in the
vicinity of $C_1$.
Circles represent  points on the coexistence lines:
open circles are for the gas-LDL coexistence,  
filled circles for the gas-HDL coexistence.
Lines are guides for the eyes.
The solid black line is the gas-HDL coexistence line.
The  red line is the gas-LDL
coexistence line. 
The solid red line is stable, while the dashed red line is 
metastable, with respect to the HDL phase.
The orange line is the LDL-HDL coexistence line.
The triangle represents the {\it triple point}.
The projection of the 
spinodal line  is represented in (a) and (b) by diamonds with dashed lines.
The spinodal line is folded in this projection, with two cusps
corresponding to the two maxima in Fig.\ref{franzese_fig1}.
Critical points occur where the coexistence lines meet these cusps.
The critical point $C_1$ is for the gas-LDL transition, and
$C_2$ is for the gas-HDL transition.
{\bf (c,d)} 
$P$-$\rho$ projections of panels (a,b). The colors, symbols and patterns of
coexistence lines, triple and critical points are the same as in panel (a,b).
Dashed blue, black and orange lines schematically
represent isotherms at the temperatures of $C_1$, $C_2$ and of the triple point,
respectively. 
Both gas-LDL and gas-HDL coexistence lines show a
local maximum, representing the estimates of $C_1$ and $C_2$, respectively. 
In all the panels, where not shown, the errors are smaller than the symbol
size.
}
\label{franzese_fig2}
\end{figure}

\begin{figure}
\mbox{ \epsfxsize=8cm  \epsffile{ 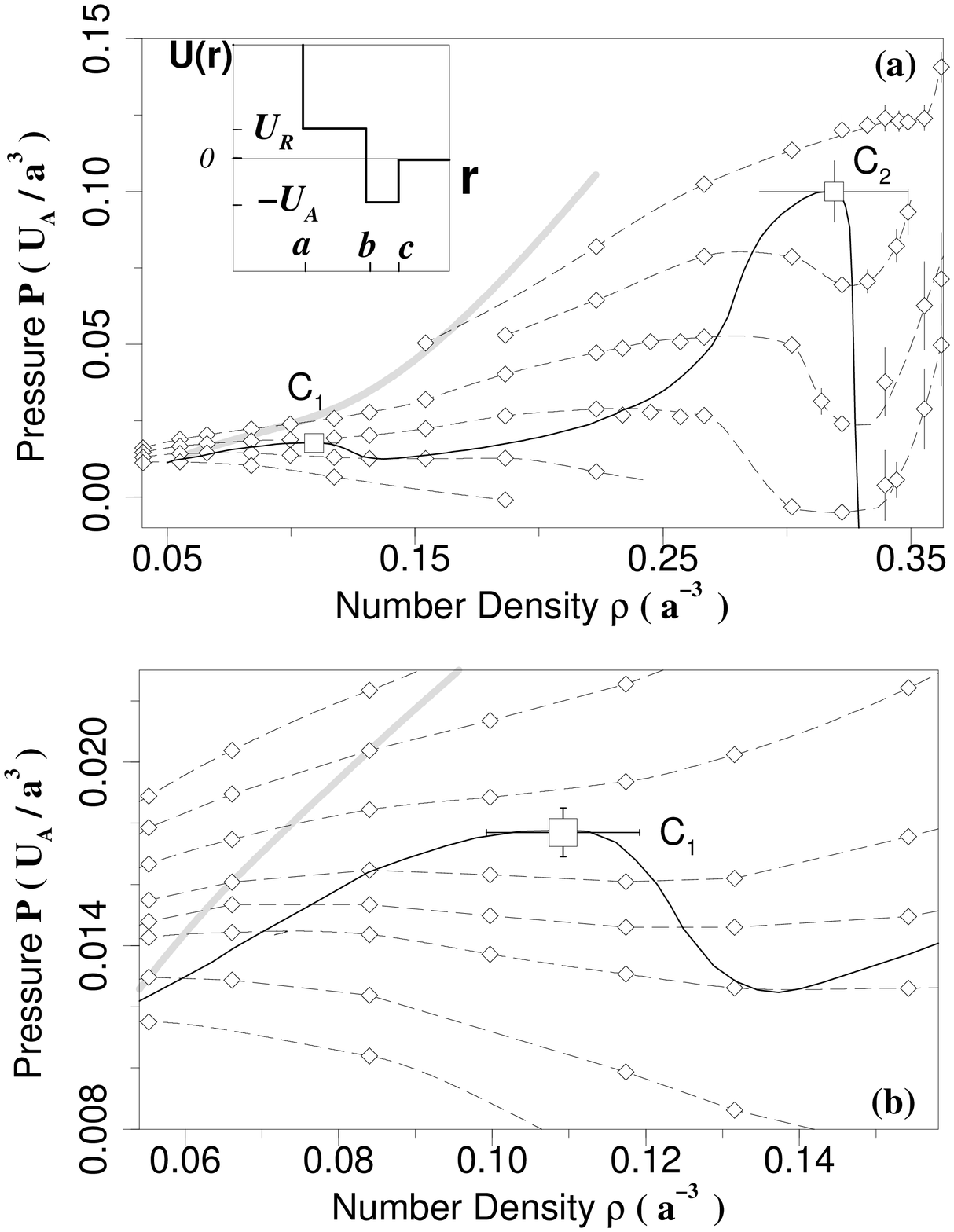 } }
\hfill{Fig.1}
\end{figure}

\begin{figure}
\mbox{ \epsfxsize=8cm  \epsffile{ 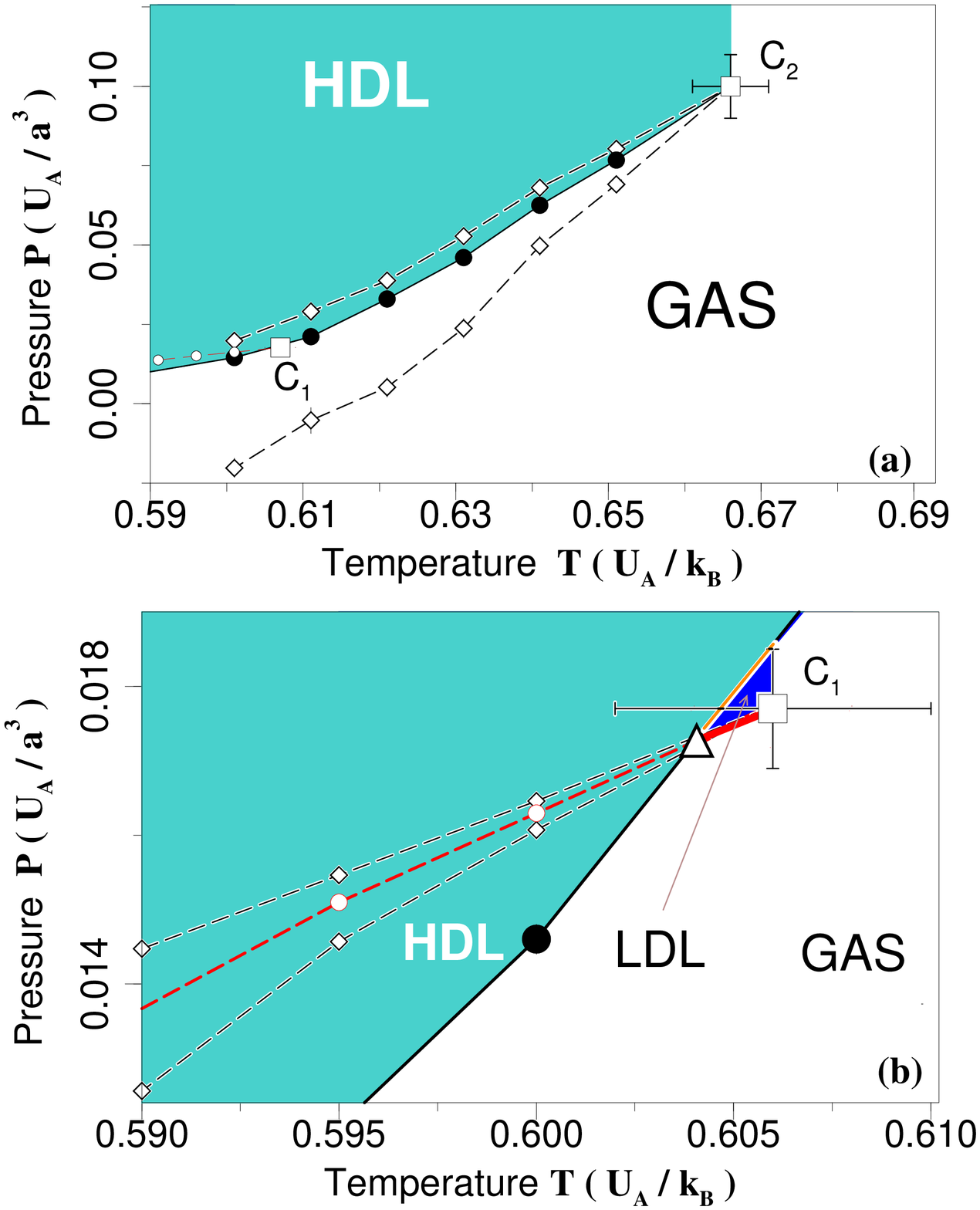 } }
\hfill{Fig.2}
\end{figure}

\begin{figure}
\mbox{ \epsfxsize=8cm  \epsffile{ 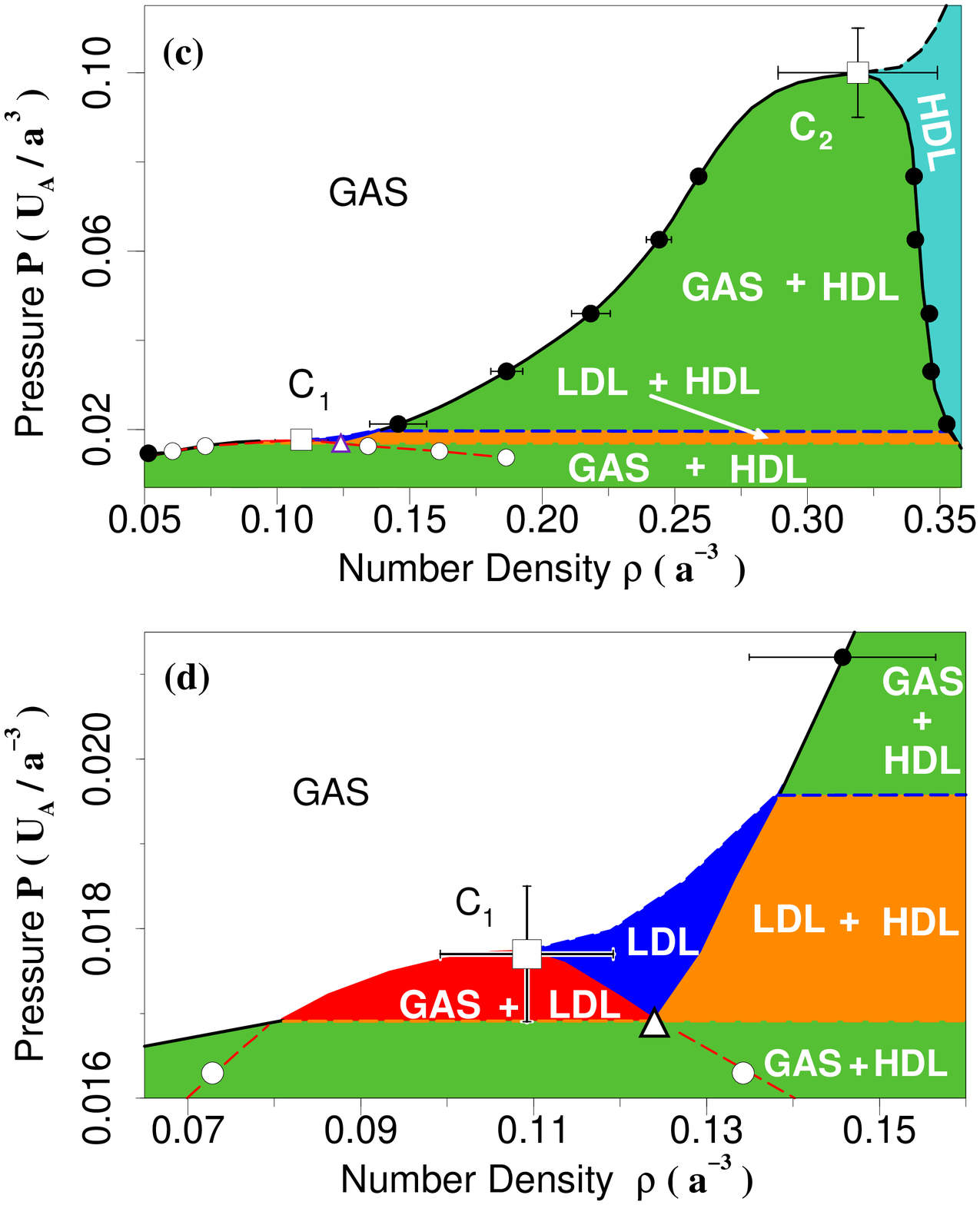 } }
\end{figure}

\end{multicols}


\bigskip

\bigskip

\bigskip


\begin{thebibliography}{99}

\bibitem{Katayama} Katayama, Y., Mizutani, T., Utsumi, W., Shimomura,
O., Yamakata, M., \& Funakoshi, K.  A first-order liquid-liquid phase
transition in phosphorus.  {\it Nature} {\bf 403}, 170--173 (2000).

\bibitem{debenedetti} Debenedetti, P.G.  {\it Metastable Liquids:
Concepts and Principles} (Princeton Univ. Press, Princeton. 1998).

\bibitem{Wilding} Wilding, M. C., McMillan, P. F., \& Navrotsky A.  The
thermodynamic nature of a phase transition in yttria-alumina
liquids. {\it J. Noncryst. Solids} (in the press). 

\bibitem{Brazhkin97} Brazhkin, V. V., Popova, S. V. \& Voloshin, R. N.
High-pressure transformations in simple melts. {\it High Pressure Res.}
{\bf 15}, 267--305 (1997).

\bibitem{Brazhkin98} Brazhkin, V. V., Gromnitskaya, E. L., Stalgorova,
O. V. \& Lyapin, A.  G.  Elastic softening of amorphous H$_2$O network
prior to the {\it HDA-LDA} transition in amorphous state. {\it Rev. High
Pressure Sci.  Tech.} {\bf 7}, 1129--1131 (1998).


\bibitem{mishima_2000} Mishima, O.  Liquid-liquid critical point in
heavy water.  {\it Phys. Rev. Lett.}  {\bf 85}, 334--336 (2000).

\bibitem{bellisent} Bellissent-Funel, M.-C.  
Evidence of a possible liquid-liquid phase transition in supercooled
water by neutron diffraction.
{\it Nuovo Cimento} {\bf 20D}, 2107-2122 (1998).

\bibitem{soper} Soper, A. K. \& Ricci, M. A.  Structures of high-density
and low-density water.  {\it Phys. Rev. Lett.}  {\bf 84}, 2881--2884
(2000).

\bibitem{Lacks} Lacks, D. J.  First-order amorphous-amorphous
transformation in silica.  {\it Phys. Rev. Lett.}  {\bf 84}, 4629--4632
(2000).


\bibitem{Thiel} van Thiel M. \& Ree F. H.  High-pressure liquid-liquid
phase change in carbon.  {\it Phys. Rev. B} {\bf 48}, 3591--3599 (1993).

\bibitem{Poole}
Poole, P. H., Sciortino, F., Essmann, U., \& Stanley, H. E.  Phase
behavior of metastable water.  {\it Nature\/} {\bf 360}, 324--328
(1992).

\bibitem{Glosli} Glosli, J. N. \& Ree, F. H.  Liquid-Liquid Phase
Transformation in Carbon.  {\it Phys. Rev. Lett.}  {\bf 82}, 4659--4662
(1999).

\bibitem{Saika-Voivod} Saika-Voivod I., Sciortino F. \& Poole P. H.
Computer simulations of liquid silica: Equation of state and
liquid-liquid phase transition.  {\it Phys. Rev. E} {\bf 63}, vol. 1 (2000) (in
press).

\bibitem{Mon} Mon, K. K., Ashcroft, M. W. \& Chester, G. V.  Core
polarization and the structure of simple metals.  {\it Phys. Rev. B}
{\bf 19}, 5103--5118 (1979).

\bibitem{hs1} Stell, G. \& Hemmer, P.C.  Phase transition due to
softness of the potential core.  {\it J. Chem. Phys.}  {\bf 56},
4274--4286 (1972).
	
\bibitem{Silbert} Silbert, M. \& Young, W.H., 
Liquid metals with structure factor shoulders.
{\it Phys. Lett.} {\bf 58A}, 469--470 (1976).

\bibitem{Levesque} 
Levesque, D. \& Weis, J.J., 
Structure factor of a system with shouldered hard sphere potential.
{\it Phys. Lett.} {\bf 60A}, 473--474 (1977).

\bibitem{Kincaid} 
Kincaid, J.M. \& Stell, G.
Structure factor of a one-dimensional shouldered hard-sphere fluid.
{\it Phys. Lett.} {\bf 65A}, 131--134 (1978).

\bibitem{Cummings} 
Cummings, P.T. \& Stell, G.
Mean spherical approximation for a model liquid metal potential.
{\it Mol. Phys.} {\bf 43}, 1267--1291 (1981).

\bibitem{Velasco} 
Velasco, E., Mederos, L., Navascu\'es, G., Hemmer, P.C., \& Stell, G.
Complex phase behavior induced by repulsive interactions.
{\it Phys. Rev. Lett.} {\bf 85}, 122--125 (2000).

\bibitem{Voronel} 
Voronel, A., Paperno, I., Rabinovich, S., \& Lapina, E.
New critical point at the vicinity of freezing temperature of K$_2$Cs.
{\it Phys. Rev. Lett.} {\bf 50}, 247--249 (1983).

\bibitem{colloids} Behrens, S. H., Christl, D. I., Emmerzael, R.,
Schurtenberger, P., \& Borkovec, M.  Charging and aggregation properties
of carboxyl latex particles: Experiments versus DLVO theory.  {\it
Langmuir} {\bf16}, 2566--2575 (2000).

\bibitem{DRB} Debenedetti, P. G., Raghavan, V. S. \& Borick, S. S.
Spinodal curve of some supercooled liquids.  {\it J. Phys. Chem.}  {\bf
95}, 4540--4551 (1991).

\bibitem{ssbs}
 Sadr-Lahijany, M.R., Scala, A., Buldyrev, S.V., \&
Stanley, H. E.  Liquid state anomalies for the Stell-Hemmer
core-softened potential.  {\it Phys. Rev. Lett.}  {\bf 81}, 4895--4898
(1998).

\bibitem{jagla} Jagla, E. A.  Core-softened potentials and the anomalous
properties of water.  {\it J. Chem. Phys.}  {\bf 111}, 8980--8986
(1999).

\bibitem{SH-G1} Stillinger, F. H. \& Head-Gordon, T.  Perturbational
view of inherent structures in water.  {\it Phys. Rev. E\/} {\bf 47},
2484--2490 (1993).

\bibitem{caccamo} Caccamo, C.  Integral equation theory description of
phase equilibria in classical fluids.  {\it Phys. Rep.}  {\bf 274},
1--105 (1996).

\bibitem{b} Berendsen, H. J. C., Postma, J. P. M., van Gunsteren, W. F.,
DiNola, A. \& Haak, J. R.  Molecular dynamics with coupling to an
external bath.  {\it J. Chem. Phys.}  {\bf 81}, 3684--3690 (1984).

\bibitem{Frenkel} 
Rein ten Wolde, P. \& Frenkel, D.
Enhancement of protein crystal nucleation by critical density
fluctuations.
{\it Science} {\bf 277}, 1975--1978 (1997). 

\bibitem{Hagen} Hagen, M. H. J., Meijer, E. J., Mooij, G. C. A. M.,
Frenkel, D., Lekkerkerker, H. N. W.  Does C-60 have a liquid-phase?
{\it Nature} {\bf 365}, 425--426 (1993).

\end{thebibliography}
\end{document}